\documentclass[12pt,a4paper,final]{revtex4}

\usepackage{sidecap}
\usepackage{ulem}
\usepackage{epsfig}
\usepackage{amsmath,amssymb,amsthm}
\usepackage{graphicx}
\usepackage{bm}
\usepackage{color,soul}

\DeclareMathAlphabet{\mathpzc}{OT1}{pzc}{m}{it}

\begin{document}

\renewcommand{\textfraction}{0.00}


\newcommand{\vAi}{{\cal A}_{i_1\cdots i_n}}
\newcommand{\vAim}{{\cal A}_{i_1\cdots i_{n-1}}}
\newcommand{\vAbi}{\bar{\cal A}^{i_1\cdots i_n}}
\newcommand{\vAbim}{\bar{\cal A}^{i_1\cdots i_{n-1}}}
\newcommand{\htS}{\hat{S}}
\newcommand{\htR}{\hat{R}}
\newcommand{\htB}{\hat{B}}
\newcommand{\htD}{\hat{D}}
\newcommand{\htV}{\hat{V}}
\newcommand{\cT}{{\cal T}}
\newcommand{\cM}{{\cal M}}
\newcommand{\cMs}{{\cal M}^*}
\newcommand{\vk}{\vec{\mathbf{k}}}
\newcommand{\bk}{\bm{k}}
\newcommand{\kt}{\bm{k}_\perp}
\newcommand{\kp}{k_\perp}
\newcommand{\km}{k_\mathrm{max}}
\newcommand{\vl}{\vec{\mathbf{l}}}
\newcommand{\bl}{\bm{l}}
\newcommand{\bK}{\bm{K}}
\newcommand{\bb}{\bm{b}}
\newcommand{\qm}{q_\mathrm{max}}
\newcommand{\vp}{\vec{\mathbf{p}}}
\newcommand{\bp}{\bm{p}}
\newcommand{\vq}{\vec{\mathbf{q}}}
\newcommand{\bq}{\bm{q}}
\newcommand{\qt}{\bm{q}_\perp}
\newcommand{\qp}{q_\perp}
\newcommand{\bQ}{\bm{Q}}
\newcommand{\vx}{\vec{\mathbf{x}}}
\newcommand{\bx}{\bm{x}}
\newcommand{\tr}{{{\rm Tr\,}}}
\newcommand{\bc}{\textcolor{blue}}

\newcommand{\beq}{\begin{equation}}
\newcommand{\eeq}[1]{\label{#1} \end{equation}}
\newcommand{\ee}{\end{equation}}
\newcommand{\bea}{\begin{eqnarray}}
\newcommand{\eea}{\end{eqnarray}}
\newcommand{\beqar}{\begin{eqnarray}}
\newcommand{\eeqar}[1]{\label{#1}\end{eqnarray}}

\newcommand{\half}{{\textstyle\frac{1}{2}}}
\newcommand{\ben}{\begin{enumerate}}
\newcommand{\een}{\end{enumerate}}
\newcommand{\bit}{\begin{itemize}}
\newcommand{\eit}{\end{itemize}}
\newcommand{\ec}{\end{center}}
\newcommand{\bra}[1]{\langle {#1}|}
\newcommand{\ket}[1]{|{#1}\rangle}
\newcommand{\norm}[2]{\langle{#1}|{#2}\rangle}
\newcommand{\brac}[3]{\langle{#1}|{#2}|{#3}\rangle}
\newcommand{\hilb}{{\cal H}}
\newcommand{\pleft}{\stackrel{\leftarrow}{\partial}}
\newcommand{\pright}{\stackrel{\rightarrow}{\partial}}

\title{DREENA-B framework: first predictions of $R_{AA}$ and $v_2$ within dynamical energy loss formalism in evolving QCD medium}

\author{Dusan Zigic}
\affiliation{Institute of Physics Belgrade, University of Belgrade, Serbia}

\author{Igor Salom}
\affiliation{Institute of Physics Belgrade, University of Belgrade, Serbia}

\author{Jussi Auvinen}
\affiliation{Institute of Physics Belgrade, University of Belgrade, Serbia}

\author{Marko Djordjevic}
\affiliation{Faculty of Biology, University of Belgrade, Serbia}

\author{Magdalena Djordjevic\footnote{E-mail: magda@ipb.ac.rs}}
\affiliation{Institute of Physics Belgrade, University of Belgrade, Serbia}

\begin{abstract} Dynamical energy loss formalism allows generating state-of-the-art suppression predictions in finite size QCD medium, employing a sophisticated model of high-$p_\perp$ parton interactions with QGP. We here report a major step of introducing medium evolution in the formalism though $1+1D$ Bjorken (``B'') expansion, while preserving all complex features of the original dynamical energy loss framework.
We use this framework to provide joint $R_{AA}$ and $v_2$ predictions, for the first time within the dynamical energy loss formalism in evolving QCD medium. The predictions are generated for a wide range of observables, i.e. for all types of probes (both light and heavy) and for all centrality regions in both $Pb+Pb$ and $Xe+Xe$ collisions at the LHC. Where experimental data are available, DREENA-B framework leads to a good joint agreement with $v_2$ and $R_{AA}$ data. Such agreement is encouraging, i.e. may lead us closer to resolving $v_2$ puzzle (difficulty of previous models to jointly explain $R_{AA}$ and $v_2$ data), though this still remains to be thoroughly tested by including state-of-the-art medium evolution within DREENA framework. While introducing medium evolution significantly changes $v_2$ predictions, $R_{AA}$ predictions remain robust and moreover in a good agreement with the experimental data; $R_{AA}$ observable is therefore suitable for calibrating parton-medium interaction model, independently from the medium evolution. Finally, for heavy flavor, we observe a strikingly similar signature of the dead-cone effect on both $R_{AA}$ and $v_2$ - we also provide a simple analytical understanding behind this result. Overall, the results presented here indicate that DREENA framework is a reliable tool for QGP tomography.

\end{abstract}

\pacs{12.38.Mh; 24.85.+p; 25.75.-q}
\maketitle

\section{Introduction}

It is by now established that Quark-gluon plasma (QGP), being a new state of matter~\cite{Collins,Baym} consisting of interacting quarks, antiquarks and gluons, is created in ultra-relativistic heavy ion collisions at the Relativistic Heavy Ion Collider (RHIC) and the Large Hadron Collider (LHC). Energy loss of rare high $p_\perp$ particles, which are created in such collisions and which transverse QGP, is considered to be an excellent probe of this form of matter~\cite{QGP1,QGP2,QGP3,QGP4}. Such energy loss is reflected through different observables, most importantly angular averaged ($R_{AA}$)~\cite{ALICE_CH_RAA,ATLAS_CH_RAA,CMS_CH_RAA,ALICE_D_RAA,RAA_ALICE2013,ALICE_D_RAA_2012,PHENIX_RAA,STAR_RAA} and angular differential ($v_2$)~\cite{ALICE_CH_v2,ATLAS_CH_v2,CMS_CH_v2,ALICE_D_v2,CMS_D_v2,ALICE_D_v2_5,ATLAS_v2,CMS_v2_2012} nuclear modification factor, which can be measured and predicted for both light and heavy flavor probes. Therefore, comparing a comprehensive set of predictions, created under the same model and parameter set, with the corresponding experimental data, allows for systematical investigation of QCD medium properties, i.e. QGP tomography.

We previously showed that the dynamical energy loss formalism~\cite{MD_PRC,DH_PRL,MD_Coll} provides an excellent tool for such tomography. In particular, we demonstrated that the formalism shows a very good agreement~\cite{MD_PLB,DDB_PLB,MD_5TeV,MD_PRL} with a wide range of $R_{AA}$ data, coming from different experiments, collision energies, probes and centralities. Recently, we also used this formalism to generate first $v_2$ predictions, within DREENA-C framework~\cite{DREENA-C}, where DREENA stands for Dynamical Radiative and Elastic ENergy loss Approach, and "C" denotes constant temperature QCD medium. These predictions were compared jointly with $R_{AA}$ and $v_2$ data, showing a very good agreement with $R_{AA}$ data, while visibly overestimating $v_2$ data. This overestimation also clearly differentiates the dynamical energy loss from other models, which systematically underestimated the $v_2$ data, leading to so called $v_2$ puzzle~\cite{v2Puzzle,Betz,Molnar_v2}. On the other hand, it is also clear that $v_2$ predictions have to be further improved - in particular $v_2$ was shown to be sensitive to medium evolution, while in DREENA-C medium evolution was introduced in the simplest form, through constant medium temperature. This problem then motivated us to introduce medium evolution in DREENA framework.

While several energy loss models already contain a sophisticated medium evolution, they employ simplified energy loss models. On the other hand, the dynamical energy loss formalism corresponds to the other "limit", where constant (mean) medium temperature was assumed, combined with a sophisticated model of parton-medium interactions, which includes:  {\it i)} QCD medium composed of dynamical (i.e. moving) scattering centers, which is contrary to the widely used static scattering centers approximation, {\it ii)} finite size QCD medium, {\it iii)} finite temperature QCD medium, modeled by generalized HTL approach~\cite{Kapusta,Le_Bellac}, naturally regularizing all infrared and ultraviolet divergencies~\cite{DG_TM,MD_PRC,DH_PRL,MD_Coll}. {\it iv)} collisional~\cite{MD_Coll} and radiative~\cite{MD_PRC} energy losses, calculated within the same theoretical framework, {\it v)} finite parton mass, making the formalism applicable to both light and heavy flavor, {\it vi)} finite magnetic~\cite{MD_MagnMass} mass and running coupling~\cite{MD_PLB}.

Note that we previously showed that all the ingredients stated above are important for accurately describing experimental data~\cite{BD_JPG}. Consequently, introducing medium evolution in the dynamical energy loss, is a major step in the model development, as all components in the model have to be preserved, and no additional simplifications should be used in the numerical procedure. In addition to developing the energy loss expressions with changing temperature, we also wanted to develop a framework that can efficiently generate a set of predictions for all types of probes and all centrality regions. That is, we think that for a model to be realistically compared with experimental data, the comparison should be done for a comprehensive set of light and heavy flavor experimental data, through the same numerical framework and the same parameter set. To implement this principle, we also had to develop a numerical framework that can efficiently (i.e. in a short time frame) generate such predictions, which will be presented in this paper.

We will start the task of introducing the medium evolution in the dynamical energy loss formalism with DREENA-B framework presented here, where "B" stands for Bjorken. In this framework, QCD medium is modeled by the ideal hydrodynamical $1+1D$ Bjorken expansion~\cite{BjorkenT}, which has a simple analytical form of temperature ($T$) dependence. This simple $T$ dependence will be used as an intermediate between constant (mean) temperature DREENA-C framework and the full evolution QGP tomography tool. While, on one hand, inclusion of Bjorken expansion in DREENA framework is a major task (having in mind complexity of our model, see above), it on the other hand significantly simplifies the numerical procedure compared to full medium evolutions. This will then allow step-by-step development of full QGP tomography framework, and assessing improvements in the predictions when, within the same theoretical framework, one is transitioning towards more complex QGP evolution models within the dynamical energy loss framework.

\section{Computational framework}

To calculate the quenched spectra of hadrons, we use the generic pQCD convolution, while the assumptions are provided in~\cite{MD_PLB}:
\begin{eqnarray}
\frac{E_f d^3\sigma}{dp_f^3} = \frac{E_i d^3\sigma(Q)}{dp^3_i}
 \otimes
{P(E_i \rightarrow E_f )}
\otimes D(Q \to H_Q) \otimes f(H_Q \to e, J/\psi), \;
\label{schem} \end{eqnarray}
where "i"  and "f", respectively, correspond to "initial" and "final", $Q$ denotes quarks and gluons. $E_i d^3\sigma(Q)/dp_i^3$ denotes the initial quark spectrum, computed at next to leading order~\cite{Vitev0912} for light and heavy partons.
$D(Q \to H_Q)$ is the fragmentation function of parton (quark or gluon)
$Q$ to hadron $H_Q$; for charged hadrons, D and B mesons we use DSS~\cite{DSS}, BCFY~\cite{BCFY} and KLP~\cite{KLP} fragmentation functions, respectively. $P(E_i \rightarrow E_f )$ is the energy loss probability, generalized to include both radiative and collisional energy loss in a realistic finite size dynamical QCD medium in which the temperature is changing, as well as running coupling, path-length  and multi-gluon fluctuations. In below expressions, running coupling is introduced according to~\cite{MD_PLB}, where  we note that temperature $T$ now changes with proper time $\tau$; the temperature dependence along the jet path is taken according to the ideal hydrodynamical $1+1D$ Bjorken expansion~\cite{BjorkenT}. Partons travel different paths in the QCD medium, which is taken into account through path length fluctuations~\cite{WHDG}. Multi-gluon fluctuations take into account that the energy loss is a distribution, and are included according to~\cite{GLV_suppress,MD_PLB} (for radiative energy loss) and~\cite{Moore:2004tg,WHDG} (for collisional energy loss).

The dynamical energy loss formalism was originally developed for constant temperature QCD medium, as described in detail in~\cite{MD_PRC,DH_PRL,MD_Coll}. We have now derived collisional and radiative energy loss expressions for the medium in which the temperature is changing along the path of the jet; detailed calculations will be presented elsewhere, while the main results are summarized below.

For collisional energy loss, we obtain the following analytical expression:
\beqar
\frac{d E_{col}}{d \tau} &=& \frac{2 C_R}{\pi \, v^2} \alpha_S (E \, T) \, \alpha_S (\mu^2_E(T)) \nonumber \\
&& \hspace*{-1.5cm}
\int_0^\infty n_{eq}(|\vec{\mathbf{k}}|, T) d |\vec{\mathbf{k}}| \;
\left( \int_0^{|\vec{\mathbf{k}}|/(1+v)} d |\vec{\mathbf{q}}|
\int_{-v |\vec{\mathbf{q}}|}^{v |\vec{\mathbf{q}}|}\; \omega d \omega \;+
\int_{|\vec{\mathbf{k}}|/(1+v)}^{|\vec{\mathbf{q}}|_{max}} d |\vec{\mathbf{q}}|
\int_{|\vec{\mathbf{q}}|-2|\vec{\mathbf{k}}| }^{v |\vec{\mathbf{q}}|}\;
\omega d \omega \; \right) \nonumber \\
&& \hspace*{-1.5cm}\left( |\Delta_L(q,T)|^2 \frac{(2 |\vec{\mathbf{k}}|+\omega)^2
- |\vec{\mathbf{q}}|^2}{2}  +
|\Delta_T(q,T)|^2 \frac{(|\vec{\mathbf{q}}|^2-\omega^2)
((2 |\vec{\mathbf{k}}|+\omega)^2+ |\vec{\mathbf{q}}|^2)}
{4 |\vec{\mathbf{q}}|^4} (v^2 |\vec{\mathbf{q}}|^2-\omega^2) \right).
\eeqar{Eel_infinite}
Here $E$ is initial jet energy, $\tau$ is the proper time, $T$ is the temperature of the medium, $\alpha_S$ is running coupling~\cite{MD_PLB} and $C_R=\frac{4}{3}$. $k$ is the 4-momentum of the incoming medium parton, $v$ is velocity of the incoming jet and $q=(\omega, \vec{\mathbf{q}})$ is the 4-momentum of the gluon. $n_{eq}(|\vec{\mathbf{k}}|, T)=\frac{N}{e^{|\vec{\mathbf{k}}|/T}-1}+\frac{N_f}{e^{|\vec{\mathbf{k}}|/T}+1}$ is the equilibrium momentum distribution~\cite{BT} at temperature $T$ including quarks and gluons ($N$ and $N_f$ are the number of colors and flavors, respectively). $\Delta_L (T)$ and $\Delta_T (T)$ are effective longitudinal and transverse gluon propagators~\cite{Gyulassy_Selikhov}:
\beq
\Delta_{L}^{-1} (T)= \vec{\mathbf{q}}^{2}+ \mu_E(T)^{2}
(1+\frac{\omega }{2|\vec{\mathbf{q}}|}
\ln |\frac{\omega -|\vec{\mathbf{q}}|}{\omega +|\vec{\mathbf{q}}|}|),
\eeq{Delta_L}
\beq
\Delta_T^{-1} (T) = \omega^2 - \vec{\mathbf{q}}^{2} - \frac{\mu_E(T)^{2}}{2} -
\frac{(\omega ^{2} - \vec{\mathbf{q}}^{2})\mu_E (T)^{2}}{2 \vec{\mathbf{q}}^{2}}
(1+\frac{\omega }{2|\vec{\mathbf{q}}|}
\ln |\frac{\omega -|\vec{\mathbf{q}}|}{\omega +|\vec{\mathbf{q}}|}|),
\eeq{Delta_T}
while the electric screening (the Debye mass) $\mu_E(T)$ can be obtained by self-consistently solving the expression~\cite{Peshier} ($n_f$ is number of the effective degrees of freedom, $\Lambda_{QCD}$ is perturbative QCD scale):
\beqar
\frac{\mu_E (T)^2}{\Lambda_{QCD}^2} \ln \left(\frac{\mu_E (T)^2}{\Lambda_{QCD}^2}\right)=\frac{1+n_f/6}{11-2/3 \, n_f} \left(\frac{4 \pi T}{\Lambda_{QCD}} \right)^2.
\eeqar{muE}

The gluon radiation spectrum takes the following form:
\beqar
\frac{dN_{\mathrm{rad}}}{dx d \tau} &=&
 \int \frac{d^2k}{\pi} \,\frac{d^2q}{\pi} \, \frac{2\, C_R  C_2(G)\, T}{x} \, \frac{\alpha_s (E \, T) \, \alpha_s (\frac{\bk^2+\chi (T)}{x}) }{\pi}\,
    \frac{\mu_E (T)^2 -\mu_M(T)^2 }{(\bq^2+\mu_M (T)^2) (\bq^2+\mu_E (T)^2)} \times\,\nonumber\\
    && \hspace*{-1cm} \times\left(1-\cos{\frac{(\bk{+}\bq)^2+\chi(T)}{x E^+} \, \tau}\right)
    \frac{(\bk{+}\bq)}{(\bk{+}\bq)^2+\chi(T)}
    \left(\frac{(\bk{+}\bq)}{(\bk{+}\bq)^2+\chi(T)}
    - \frac{\bk}{\bk^2+\chi(T)}
    \right),
\eeqar{DeltaNDynTau}
where $C_2(G)=3$ and $\mu_M (T)$ is magnetic screening. $\bk$ and $\bq$ are transverse momenta of radiated and exchanged (virtual) gluon, respectively.
$\chi (T) \equiv M^2 x^2 + m_E (T)^2/2$, where $x$ is the longitudinal
momentum fraction of the jet carried away by the emitted gluon, $M$ is the mass of the quark of gluon jet and $m_g (T)= \mu_E(T)/\sqrt(2)$ is effective gluon mass in finite temperature QCD medium~\cite{DG_TM}. We also recently abolished the soft-gluon approximation~\cite{Blagojevic:2018nve}, for which we however showed that it does not significantly affect the model results; consequently, this improvement is not included in DREENA-B, but can be straightforwardly implemented in the future DREENA developments, if needed.

Note that, as a result of introducing medium evolution, we got that the dynamical energy loss formalism now explicitly contains changing temperature in the energy loss expression. This is contrary to most of the other models, in which temperature evolution is introduced indirectly, through $\hat{q}$ or $\frac{dNg}{dy}$ (see~\cite{JET} and references therein). This then makes the dynamical energy loss a natural framework to incorporate diverse temperature profiles as a starting point for QGP tomography. As a first (major) step, we will below numerically implement this possibility through Bjorken $1+1D$ expansion~\cite{BjorkenT}.

Regarding the numerical procedure, computation efficiency of the algorithm implemented in DREENA-C framework~\cite{DREENA-C} was already two orders of magnitude higher with respect to the basic (unoptimized) brute-force approach applied in~\cite{MD_PLB}. However, straightforward adaptation of the DREENA-C code to the case of the Bjorken type evolving medium was not sufficient. This was dominantly due to additional integration over proper time $\tau$, which increased the calculation time for more than two orders of magnitude. The computation of e.g. the radiative energy losses alone, for a single probe, took around 10 hours on the available computer resources (a high performance workstation). Taking into account that it requires $\sim 10^3$ such runs to produce the results presented in this paper, it is evident that a substantial computational speedup was necessary.

The main algorithmic tool that we used to optimize the calculation was a combination of sampling and tabulating various intermediate computation values and their subsequent interpolation. We used nonuniform adaptive grids of the sampling points, denser in the parts of the parameter volume where the sampled function changed rapidly. Similarly, the parameters used for the numerical integration (the number of Quasi Monte Carlo sampling points and the required accuracy) were also suitably varied throughout the parameter space. Finally, while the computation in DREENA-C was performed in a software for symbolic computation, the new algorithm was redeveloped in C programming language. The combined effect of all these improvements was a computational speedup of almost three orders of magnitude, which was a necessary prerequisite for both current practical applicability and future developments of DREENA framework.

Regarding the parameters, we implement Bjorken $1+1D$ expansion~\cite{BjorkenT}, with commonly used $\tau_0=0.6$~fm~\cite{tau0KH,tau0BMB}, and initial temperatures for different centralities calculated according to $T_0 \sim (dN_{ch}/dy/ A_\perp)^{1/3}$~\cite{T0ref}, where $dN_{ch}/dy$ is charged multiplicity and $A_\perp$ is overlap area for specific collision system and centrality. We use this equation, starting from $T_0=500$~MeV in $5.02$~TeV $Pb+Pb$ most central collisions at the LHC, which is estimated based on average medium temperature of 348~MeV in these collisions, and QCD transition temperature of $T_c \approx 150$~MeV~\cite{Tcritical}. Note that the average medium temperature of 348~MeV in most central $5.02$~TeV $Pb+Pb$ collisions comes from~\cite{DDB_PLB} the effective temperature ($T_{eff}$) of 304 MeV for 0-40$\%$ centrality $2.76$ TeV Pb+Pb collisions at the LHC~\cite{ALICE_T} experiments (as extracted by ALICE). Once $T_0$s for most central $Pb+Pb$ collisions is fixed, $T_0$ for both different centralities and different collision systems ($Xe+Xe$ and $Pb+Pb$) are obtained from the expression above.

Other parameters used in the calculation remain the same as in DREENA-C~\cite{DREENA-C}. In particular, the path-length distributions for both $Xe+Xe$ and $Pb+Pb$ are calculated following the procedure described in~\cite{Dainese}, with an additional hard sphere restriction $r < R_A$ in the Woods-Saxon nuclear density distribution to regulate the path lengths in the peripheral collisions. Note that the path-length distributions for $Pb+Pb$ are explicitly provided in~\cite{DREENA-C}; we have also checked that, for each centrality, our obtained eccentricities remain within the standard deviation of the corresponding Glauber Monte Carlo results~\cite{Ecc} (results not shown). For $Xe+Xe$, it is straightforward to show that $Xe+Xe$ and $Pb+Pb$  distributions are the same up to recalling factor ($A^{1/3}$, where $A$ is atomic number), as we discussed in~\cite{Djordjevic:2018ita}. Furthermore, the path-length distributions correspond to geometric quantity, and are therefore the same for all types of partons (light and heavy). For QGP, we take $\Lambda_{QCD}=0.2$~GeV and $n_f{\,=\,}3$. As noted above, temperature dependent Debye mass $\mu_E (T)$  is obtained from~\cite{Peshier}. For  light quarks and gluons, we, respectively, assume that their effective masses are $M{\,\approx\,}\mu_E (T)/\sqrt{6}$ and $m_g\approx \mu_E (T)/\sqrt{2}$~\cite{DG_TM}. The charm and bottom masses are $M{\,=\,}1.2$\,GeV  and $M{\,=\,}4.75$\,GeV, respectively. Magnetic to electric mass ratio is extracted from non-perturbative calculations~\cite{Maezawa,Nakamura}, leading to $0.4 < \mu_M/\mu_E < 0.6$ - this range of screening masses lead to presented uncertainty in the predictions. We note that no fitting parameters are used in the calculations, that is, all the parameters correspond to standard literature values.

\section{Results and discussion}
\begin{figure}
\epsfig{file=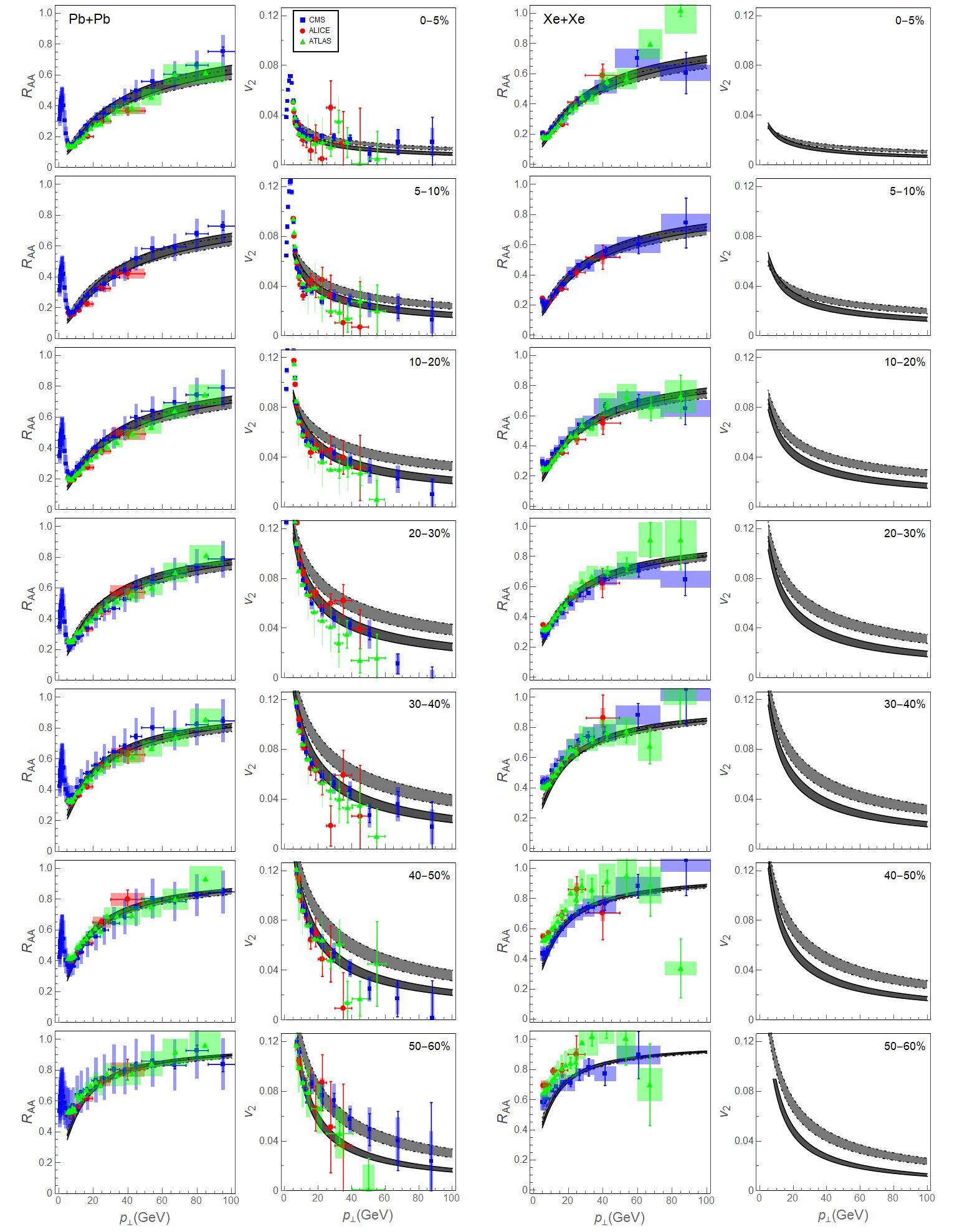,width=6.2in,height=7.5in,clip=5,angle=0}
\vspace*{-0.8cm}
\caption{ {\it First column:} $R_{AA}$ {\it vs.} $p_\perp$ predictions are compared with $5.02$~TeV $Pb+Pb$ ALICE~\cite{ALICE_CH_RAA}, ATLAS~\cite{ATLAS_CH_RAA} and CMS~\cite{CMS_CH_RAA} $h^\pm$ experimental data.  {\it Second column:} $v_2$ {\it vs.} $p_\perp$ predictions are compared with $5.02$~TeV $Pb+Pb$ ALICE~\cite{ALICE_CH_v2}, ATLAS~\cite{ATLAS_CH_v2} and CMS~\cite{CMS_CH_v2} data. {\it Third column:} $R_{AA}$ {\it vs.} $p_\perp$ predictions are compared with $5.44$~TeV $Xe+Xe$ ALICE~\cite{ALICE_XeXe}, ATLAS~\cite{ATLAS_XeXe} and CMS~\cite{CMS_XeXe} preliminary data.  {\it Fourth column:} $v_2$ {\it vs.} $p_\perp$ predictions are shown for $5.44$~TeV $Xe+Xe$ collisions. Rows 1-7 correspond to $0-5\%$, $5-10\%$, $10-20\%$,..., $50-60\%$ centrality regions. ALICE, ATLAS and CMS data are respectively represented by red circles, green triangles and blue squares. Full and dashed curves correspond, respectively, to the predictions obtained with DREENA-B and DREENA-C frameworks.  In each panel, the upper (lower) boundary of each gray band corresponds to $\mu_M/\mu_E =0.6$ ($\mu_M/\mu_E =0.4$).  }
\label{CH_RAA_v2}
\end{figure}

\begin{figure*}
\epsfig{file=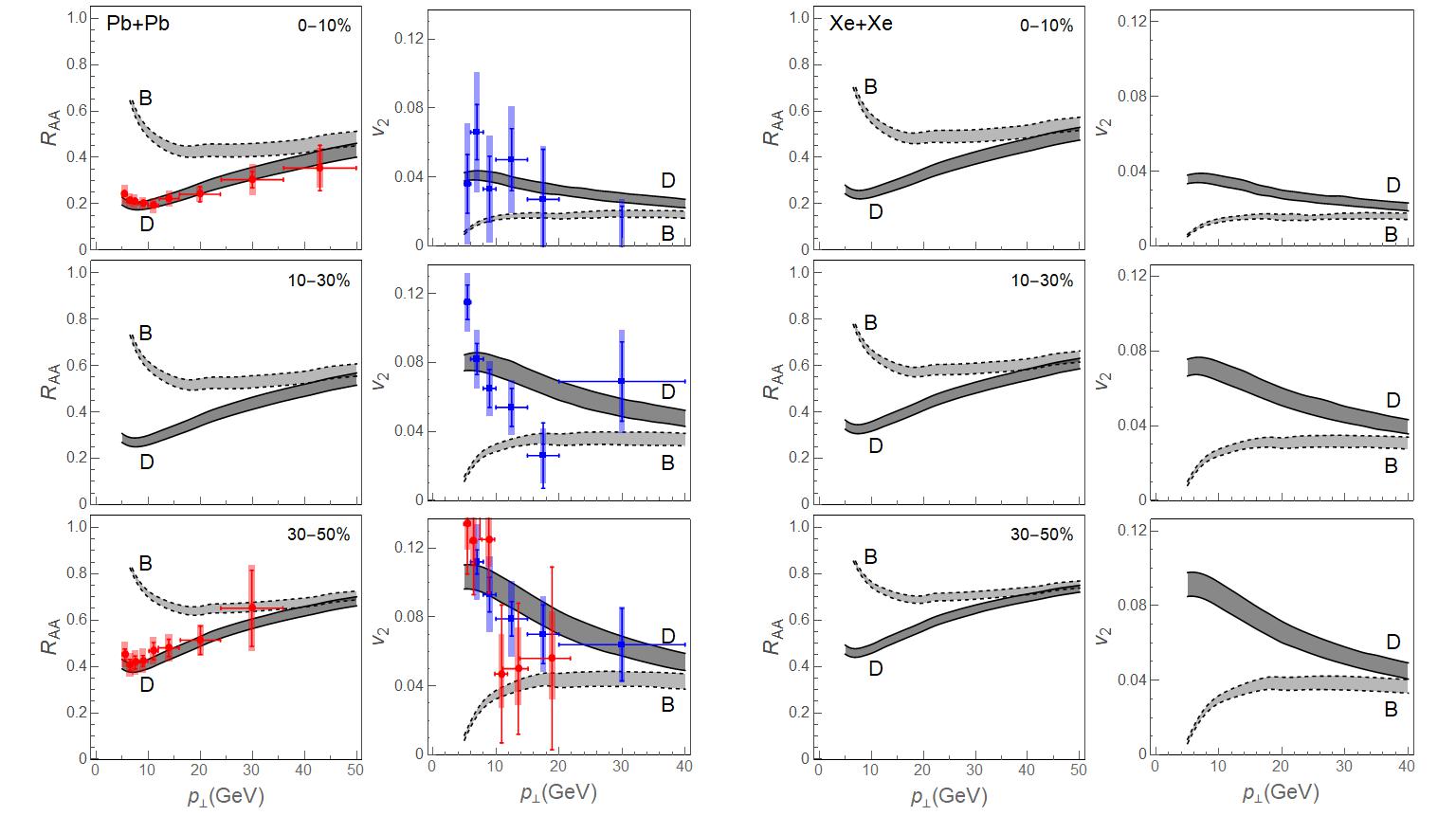,width=6.9in,height=3.9in,clip=5,angle=0}
\vspace*{-0.2cm}
\caption{  {\it First column:} Theoretical predictions for D and B meson $R_{AA}$ {\it vs.} $p_\perp$ are compared with the available $5.02$~TeV $Pb+Pb$ ALICE~\cite{ALICE_D_RAA} (red circles) D meson experimental data. {\it Second column:} $v_2$ {\it vs.} $p_\perp$ predictions are compared with $5.02$~TeV $Pb+Pb$ ALICE~\cite{ALICE_D_v2_5} (red circles) and CMS~\cite{CMS_D_v2} (blue squares) D meson experimental data. {\it Third and fourth column:}  Heavy flavor $R_{AA}$ and $v_2$ {\it vs.} $p_\perp$ predictions are, respectively, provided for $5.44$~TeV $Xe+Xe$ collisions at the LHC.  First to third row, respectively, correspond to $0-10\%$, $10-30\%$ and $30-50\%$ centrality regions. On each panel, the upper (lower) boundary of each gray band corresponds to $\mu_M/\mu_E =0.6$ ($\mu_M/\mu_E =0.4$). }
\label{DB_RAA_v2_B}
\end{figure*}

In this section, we will present joint $R_{AA}$ and $v_2$ predictions for light (charged hadrons) and heavy (D and B mesons) flavor in $Pb+Pb$  and $Xe+Xe$ collisions at the LHC, obtained by DREENA-B framework. Based on the path-length distributions from Figure~1 in~\cite{DREENA-C}, we will, in Figures~\ref{CH_RAA_v2} to~\ref{DB_RAA_v2_B}, show $R_{AA}$ and $v_2$ predictions for light and heavy flavor, in $5.02$~TeV $Pb+Pb$ and $5.44$~TeV $Xe+Xe$ collisions, at different centralities. We start by presenting charged hadrons predictions, where $R_{AA}$ data are available for both $Pb+Pb$ and $Xe+Xe$, while $v_2$ data exist for $Pb+Pb$ collisions. Comparison of our joint predictions with experimental data is shown in Figure~\ref{CH_RAA_v2}, where $1^{st}$ and $2^{nd}$ columns correspond, respectively, to $R_{AA}$ and $v_2$ predictions at $Pb+Pb$, while $3^{rd}$ and $4^{th}$ columns present equivalent predictions/data for $Xe+Xe$ collisions at the LHC. From this figure, we see that DREENA-B is able to well explain joint $R_{AA}$ and $v_2$ predictions. For $5.44$~TeV $Xe+Xe$ collisions at the LHC, we observe good agreement of our predictions with preliminary $R_{AA}$ data from ALICE, ATLAS and CMS data (where we note that these predictions were generated, and posted on arXiv, before the data became available), except for high centrality regions, where our predictions do not agree with ALICE (and partially with ATLAS) data; however, note that in these regions ALICE, ATLAS and CMS data also do not agree with eachother.

Furthermore, comparison of predictions obtained with DREENA-B and DREENA-C frameworks in Fig.~\ref{CH_RAA_v2}, allow directly assessing the importance of inclusion of medium evolution on different observables, as the main difference between these two frameworks is that DREENA-B contains Bjorken evolution, while DREENA-C accounts for evolution in simplest form (through constant mean temperature). We see that inclusion of Bjorken evolution has negligible effect on $R_{AA}$, while significant effect on $v_2$. That is, it keeps $R_{AA}$ almost unchanged, while significantly decreasing $v_2$. Consequently, small effect on $R_{AA}$, supports the fact that $R_{AA}$ is weekly sensitive to medium evolution, making $R_{AA}$ an excellent probe of jet-medium interactions in QGP; i.e. in QGP tomography, $R_{AA}$ can be used to calibrate parton medium interaction models.  On the other hand, medium evolution has significant influence on $v_2$ predictions, in line with previous conclusions~\cite{Thorsten,Molnar}; this sensitivity makes $v_2$ an ideal probe to constrain QGP medium parameters also from the point of high $p_\perp$ measurements (in addition to constraining them from low $p_\perp$ predictions and data).

In Figure~\ref{DB_RAA_v2_B}, we provide joint predictions for D and B meson $R_{AA}$ (left panel) and $v_2$ (right panel) predictions for both $5.02$~TeV $Pb+Pb$ and $5.44$~TeV $Xe+Xe$ collisions at the LHC. Predictions are compared with the available experimental data. For D mesons, we again observe good joint agreement with the available $R_{AA}$ and $v_2$ data. For B mesons (where the experimental data are yet to become available), we predict notably large suppression (see also~\cite{MD_PLB,DBZ}), which is consistent with non-prompt $J/\Psi$ $R_{AA}$ measurements~\cite{CMS_JPsi} (indirect probe od b quark suppression). Additionally, we predict non-zero $v_2$ for higher centrality regions. This does not necessarily mean that heavy B meson flows, since we here show predictions for high $p_\perp$, and flow is inherently connected with {\it low} $p_\perp$ $v_2$. On the other hand, high $p_\perp$ $v_2$ is connected with the difference in the B meson suppression for different (in-plane and out-of-plane) directions,  leading to our predictions of non zero $v_2$ for {\it high} $p_\perp$ B mesons. Additionally, by comparing D and B meson $v_2$s in Fig.~\ref{DB_RAA_v2_B}, we observe that their difference is large and that it qualitatively exhibits the same dependence on $p_\perp$ as $R_{AA}$. This $v_2$ comparison therefore presents additional important prediction of the heavy flavor dead-cone effect in QGP, where a strikingly similar signature of this effect is observed for $R_{AA}$ and $v_2$.

The predicted similarity between $R_{AA}$ and $v_2$ dead-cone effects can be analytically understood by using simple scaling arguments. Fractional energy loss can be estimated as~\cite{DREENA-C}
\begin{eqnarray}
\Delta E/E \sim \eta T^a L^b,
\label{ElossEstimate} \end{eqnarray}
where $a, b$ are proportionality factors, $T$ and $L$ are, respectively, the average temperature of the medium and the average path-length traversed by the jet. $\eta$ is a proportionality factor that depends on initial jet mass $M$ and transverse momentum $p_\perp$.

Under the assumption of small fractional energy loss, we can make the following estimate~\cite{DREENA-C}:
\begin{eqnarray}
R_{AA} &\approx& 1-\xi (M,p_\perp) T^a L^b, \nonumber \\
v_2 &\approx& \xi (M,p_\perp) \frac{(T^a L^{b-1} \Delta L - T^{a-1} L^{b} \Delta T)}{2},
\label{Raav2Estimate}
\end{eqnarray}
where $\Delta L$ and $\Delta T$ are, respectively, changes in average path-lengths and average temperatures along out-of-plane and in-plane directions. $\xi = (n-2) \eta/2$, where $n$ is the steepness of the initial momentum distribution function.

The difference between $R_{AA}$ and $v_2$ for $D$ and $B$ mesons then becomes:
\begin{eqnarray}
R^B_{AA}-R^D_{AA} &\approx& (\xi (M_c,p_\perp)-\xi (M_b,p_\perp))\,  T^a L^b, \nonumber \\
v^D_2-v^B_2 &\approx& \left(\xi (M_c,p_\perp)-\xi (M_b,p_\perp)\right) \, \frac{(T^a L^{b-1} \Delta L - T^{a-1} L^{b} \Delta T)}{2},
\label{Raav2deadcone} \end{eqnarray}
where $M_c$ and $M_b$ are charm and bottom quark masses respectively. From Eq.~\ref{Raav2deadcone}, we see the same mass dependent prefactor for both $R_{AA}$ and $v_2$ comparison, intuitively explaining our predicted dead-cone effect similarity for high-$p_\perp$ $R_{AA}$ and $v_2$.

\section{Summary}

Overall, we see that comprehensive joint $R_{AA}$ and $v_2$ predictions, obtained with our DREENA-B framework, lead to a good agreement with all available light and heavy flavor data. This is, to our knowledge, the first study to provide such comprehensive predictions for high $p_\perp$ observables. In the context of $v_2$ puzzle, this study presents a significant development, as the other models were not able to achieve this agreement without introducing new phenomena~\cite{CUJET3}. However, for more definite conclusions, the inclusion of more complex QGP evolution within DREENA framework is needed, which is our main ongoing - but highly non-trivial - task, due to the complexity of underlying energy loss formalism.

As an outlook,  for $Xe+Xe$, we also showed an extensive set of predictions for both $R_{AA}$ and $v_2$, for different flavors and centralities, to be compared with the upcoming experimental data.  Reasonable agreement with these data would present a strong argument that the dynamical energy loss formalism can provide a reliable tool for precision QGP tomography. Moreover, such comparison between predictions and experimental data can also confirm interesting new patterns in suppression data, such as our prediction of strikingly similar signature of the dead-cone effect between $R_{AA}$ and $v_2$ data.

{\em Acknowledgments:}
We thank  Bojana Blagojevic and Pasi Huovinen for useful discussions. We thank ALICE, ATLAS and CMS Collaborations for providing the shown data. This work is supported by the European Research Council, grant ERC-2016-COG: 725741, and by  the Ministry of Science and Technological
Development of the Republic of Serbia, under project numbers ON171004, ON173052 and ON171031.

\end{document}